%% file: chil-manuscript.tex
\documentclass[pmlr,twocolumn,10pt]{jmlr} 
\usepackage{xspace}
\usepackage{afterpage}
\usepackage{stfloats}
\usepackage{layouts}
\usepackage{float}
\usepackage{tabularx}
\restylefloat{table}
\newcommand*\audio{\vcenter{\hbox{\includegraphics[height=\baselineskip]{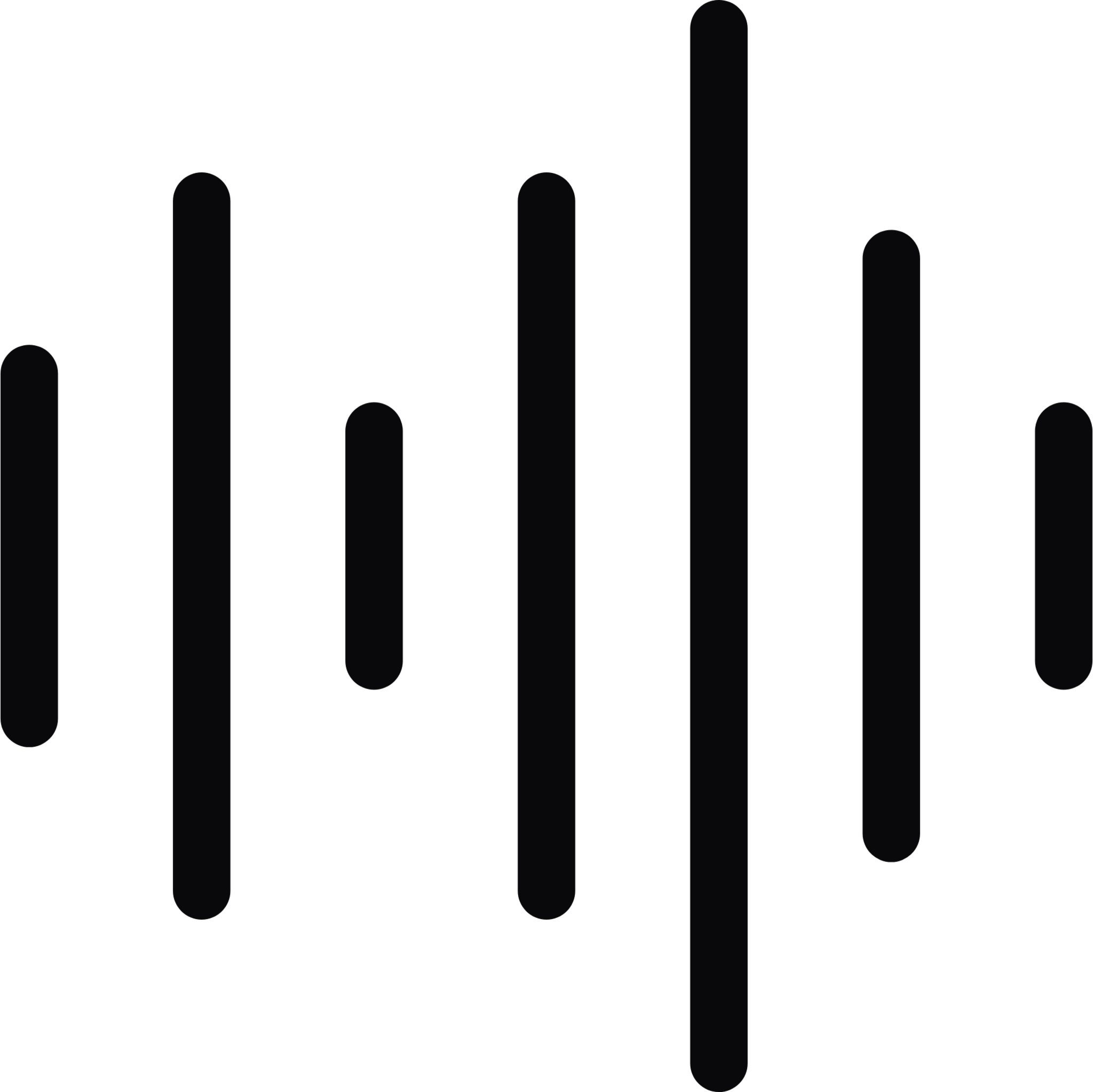}}}}
\newcommand{\datasetnihss}{Dataset NIHSS\xspace}
\newcommand{\datasetvowel}{Dataset vowel\xspace}
\input{math_commands}

\usepackage{booktabs}
\usepackage{siunitx}
\usepackage{tikz}
\usepackage{booktabs,lmodern}
\usepackage{overpic}
\usepackage{graphicx}
\usepackage{xcolor, colortbl}
\usepackage[export]{adjustbox}
\usepackage{multirow}
\usepackage{tcolorbox}
\usepackage{arydshln}
\usepackage{mathtools}
\usepackage{caption}
\definecolor{tabfirst}{rgb}{1, 0.7, 0.7} 
\definecolor{tabsecond}{rgb}{1, 0.85, 0.7} 
\definecolor{tabthird}{rgb}{1, 1, 0.7} 

\makeatletter
\newif\if@notables
\relax
\newcommand*{\ifnotables}{%
  \if@notables
    \expandafter\@firstoftwo
  \else
    \expandafter\@secondoftwo
  \fi
}
\makeatother

\usepackage{etoolbox}

\newcolumntype{L}{>{\raggedright\arraybackslash}X}
\newcolumntype{R}{>{\raggedleft\arraybackslash}X}

\usepackage[switch]{lineno}

\newcommand{\equal}[1]{{\hypersetup{linkcolor=black}\thanks{#1}}}

\theorembodyfont{\upshape}
\theoremheaderfont{\scshape}
\theorempostheader{:}
\theoremsep{\newline}

\jmlrvolume{LEAVE UNSET}
\jmlryear{2024}
\jmlrsubmitted{LEAVE UNSET}
\jmlrpublished{LEAVE UNSET}
\jmlrworkshop{Conference on Health, Inference, and Learning (CHIL) 2024} 

\title[Analysis of Audio Classifier Performance in Clinical Settings with Limited Data]{Tuning In $\audio$\ : Analysis of Audio Classifier Performance in Clinical Settings with Limited Data}

\author{%
 \Name{Hamza Mahdi}\equal{These authors share first-authorship and contributed equally}$^{1,2,3,4,5}$\Email{hmahdi2026@meds.uwo.ca} \\
 \Name{Eptehal Nashnoush}\footnotemark[1]$^{1,2,3,4}$\Email{e.nashnoush@mail.utoronto.ca}\\
 \Name{Rami Saab}\thanks{Core Contributors}$^{1,2,3,4}$\Email{rami.saab@utoronto.ca}\\
 \Name{Arjun Balachandar}\footnotemark[2]$^{1,2,3,4}$\Email{arjun.balachandar@mail.utoronto.ca}\\ 
\Name{Rishit Dagli}\footnotemark[2]$^6$\Email{rishit@cs.toronto.edu}\\
 \Name{Lucas X. Perri}$^{1,2,3,4}$\Email{lucas.perri@outlook.com}\\
 \Name{Houman Khosravani}\footnotemark[2]$^{1,2,3,4}$\Email{h.khosravani@utoronto.ca}
 \AND
 $^1$ \addr Temerty Centre for Artificial Intelligence Research and Education in Medicine, University of Toronto, Canada\\
 $^2$ \addr Hurvitz Brain Sciences Program, Toronto, Canada\\
 $^3$ \addr Division of Neurology, Department of Medicine, University of Toronto, Canada\\
 $^4$ \addr Sunnybrook Research Institute, Toronto, Canada\\
 $^5$ \addr Western University, Canada\\
 $^6$ \addr Department of Computer Science, University of Toronto, Canada
 }

\begin{document}
\maketitle

\begin{abstract}
This study assesses deep learning models for audio classification in a clinical setting with the constraint of small datasets reflecting real-world prospective data collection. We analyze CNNs, including DenseNet and ConvNeXt, alongside transformer models like ViT, SWIN, and AST, and compare them against pre-trained audio models such as YAMNet and VGGish. Our method highlights the benefits of pretraining on large datasets before fine-tuning on specific clinical data. We prospectively collected two first-of-its-kind patient audio datasets from stroke patients. We investigated various preprocessing techniques, finding that RGB and grayscale spectrogram transformations affect model performance differently based on the priors they learn from pretraining. Our findings indicate CNNs can match or exceed transformer models in small dataset contexts, with DenseNet-Contrastive and AST models showing notable performance. This study highlights the significance of incremental marginal gains through model selection, pre-training, and preprocessing in sound classification; this offers valuable insights for clinical diagnostics that rely on audio classification.  

\end{abstract}
\paragraph*{Data and Code Availability}
In this research, in addition to patient data, we utilized publicly available datasets such as ImageNet~\citep{deng2009imagenet}, AudioSet~\citep{gemmeke2017audio}, US8K~\citep{Salamon:UrbanSound:ACMMM:14} and ESC50~\citep{piczak2015dataset}, alongside a proprietary clinical dataset. Due to confidentiality, patient privacy regulations, and local research ethics board (REB) constraints, the clinical portion of the dataset cannot be shared at this time. We are committed to open-source all code on Github post-review. We are working with our local data sharing hub at our university institute to make the clinical data available in the near future through an REB amendment and the framework the institute has in place for such purposes.

\paragraph*{Institutional Review Board (IRB)}
The study involving human subjects received approval by the Research Ethics Board of our local hospital and institution. Throughout, we ensured compliance with local institution guidelines. All individuals or SDMs who participated in the study provided their written consent.

\section{Introduction}
\label{sec:intro}

Auditory biomarkers have been widely incorporated as the first line of assessment in medical applications; especially nonspeech and nonsemantic sounds have been used for decades to detect respiratory problems~\citep{Pahar_2021}. Modern tools for collecting and analyzing audio data have revolutionized the diagnosis of common symptoms like coughing, making voice analysis a critical first step in the diagnostic process.~\citep{larson2012validation, 6091487}. Furthermore, clinical problems ranging from continuous health monitoring~\citep{9486929}, stroke~\citep{masa2023paper}, psychiatric conditions~\citep{moedomo2012breath, fagherazzi2021voice}, neurodegenerative diseases~\citep{fagherazzi2021voice, VIZZA201945} to cardiac applications~\citep{dwivedi2018algorithms, emmanuel2012review} and lung pathology detection~\citep{gavriely1994respiratory, sello2008respiratory, aykanat2017classification, s20041214} among others have adopted non-speech health acoustic data as an important biomarker. However, understanding the clinical effect of different preprocessing and modeling techniques for the audio domain is a longstanding challenge.


The domain of modeling auditory data borrows heavily from the advances made in sequential and vision models. Such auditory models have greatly progressed through the
development of large-scale unsupervised pre-training for audio encoders~\citep{NEURIPS2020_92d1e1eb,9383459}. This transposition is usually facilitated by the conversion of audio signals into log-mel spectrograms or superlets, which are then analyzed using algorithms originally designed for vision or sequential data~\citep{pmlr-v202-radford23a,8553106}. Such approaches for modeling auditory data leverage the significant progress made in the vision and sequence modalities.

Among the techniques for transforming audio into such formats for learning representations from, log-mel spectrograms and superlets~\citep{moca2021time} represent two leading methodologies. Log-mel spectrograms are widely recognized for their ability to approximate the human auditory system's response to sound. This method converts audio signals into a spectrogram using the Mel scale. 
This results in a compact, yet effective representation of sound, which highlights the elements most relevant to human auditory perception and have been widely used for acoustic problems~\citep{bock2011enhanced, pmlr-v202-radford23a, 8553106}. Superlets~\citep{moca2021time}, include a set of wavelets that are iteratively applied across different cycles with a specific central frequency, potentially capturing more nuanced information within complex audio signals. We were particularly interested in the comparative efficacy of these methods for downstream clinical applications in neurology and beyond, which remains a subject of ongoing research.

Audio data is increasingly used as a biomarker in clinical settings for disease classification and assessment, and our aim was to expand the range of possibilities for both analysis and characterization of changes using different modeling techniques and at different stages of processing. Our analysis focuses on clinical data, in a neurologic setting but with broader applications, and in comparing different model approaches. Using data sets first-of-its-kind prospectively collected, \datasetnihss and \datasetvowel aim to expand the existing area of research in the context of disease state classification when starting with limited real-world data. This opens the way for use in other clinical settings (neurologic and beyond), where audio data can be used as a disease biomarker, and in settings where limited data is available. This also includes clinical settings and datasets where data is limited to not only collection ability but also rare diseases where data scarcity is an intrinsic factor.

\paragraph{Key Contributions.} 

Acoustic-based clinical diagnosis (or prognosis) has gained popularity in medical applications leading the way to consider audio data as a biomarker in disease classification, risk prediction, and monitoring. However, the impact of modeling decisions on these medical tasks remains largely unexplored, with one variable being limited datasets. There are also implications for rare diseases that intrinsically have this limitation. In this work, we focused on stroke as a neurovascular disease process and utilized speech as a biomarker and surrogate for swallowing difficulty (dysphagia). We evaluated training health acoustic models with different preprocessing techniques—mel RGB, log-mel mono, and superlet—affects clinical data representation and classifiction based on a defined clinical outcome state of dysphagia.
\begin{itemize}

    \item Introduced \datasetnihss: A novel data set that captures continuous speech, sentences and words based on the National Institutes of Health Stroke Scale (NIHSS)~\citep{kwah2014national}, an internationally established neurologic assessment scale for stroke emergencies.
    \item Introduced \datasetvowel: A unique dataset of sustained vowel sounds from patients, which further aids in the analysis of swallowing disorders.
    \item Analyzed Model Training Impact: Evaluated how training health acoustic models with different preprocessing techniques, mel RGB, log-mel mono, and superlet, affects clinical data representation and classifiction based on a defined clinical outcome state of dysphagia.
\end{itemize}



\section{Related Work}

Categorizing acoustic data is a problem that has been well explored throughout the years, and many deep learning-based methods have recently performed very well in classifying acoustic data, which has led to exploration of application in clinical settings. Often directly analyzing raw audio leads to improper learned representations, and acoustic data needs to be preprocessed first. We provide a brief overview of acoustic classification and acoustic event detection approaches. We are particularly interested in analyzing the downstream clinical effects of such approaches. We also provide a brief overview of the analysis done previously in these areas.

\subsection{Audio Classification}

Audio event detection and classification historically relied on simple representations of the underlying audio to transform the audio based on dynamic time warping (DTW), which allowed algorithms trained on these representations to measure spectral variability~\cite{1163055, salvador2007toward}. Following this Hidden Markov Models (HMM) started gaining popularity for discrete speech and soon became the dominant technique for all audio-based applications and also outperformed early neural approaches~\citep{6770768, raphael1999automatic}. Following this, the task of audio classification was mainly addressed by features such as Mel-frequency cepstrum coefficients (MFCC) and classifiers based on Gaussian Mixture Models (GMM)~\citep{nilsson2002gaussian}, Hidden Markov Models (HMM)~\citep{6770768, raphael1999automatic}, Nonnegative matrix factorization (NMF)~\citep{holzapfel2008musical, ozerov2009multichannel} or support vector machine (SVM)~\citep{dhanalakshmi2009classification}. Soon these models transitioned to a discriminative training strategy~\citep{hermansky2000tandem}, which led to weighted finite-state transducers (WFSTs) becoming increasingly common and Restricted Boltzmann Machines (RBMs) becoming the first popular neural component of acoustic models. Following this,~\citet{6163899} led to neural architectures becoming the dominant approach for modeling audio after demonstrating 30\% RER on the Switchboard benchmark.

Modern neural-network-based models for acoustic tasks have demonstrated significant performance increases over previous approaches with ConvNets~\citep{hershey2017cnn, schmid2023efficient, gong2021psla}, RNNs~\citep{phan2017audio, gimeno2020multiclass}, Transformers~\citep{koutini2021efficient, jaegle2021perceiver, chen2022beats} often trained with Self-supervised learning~\citep{gong2022contrastive, georgescu2023audiovisual, huang2022masked}, and for some tasks Diffusion~\citep{kong2020diffwave, lee2021nu} models.

The gains offered by early deep neural networks (DNN), HMM, and hybrid models could be attributed mainly to the wider frame windows used as inputs. Although these features are valuable, they are highly correlated, and neural networks become prominent for this task by building specialized acoustic architectures as opposed to building language models over acoustic frames like most of the early neural approaches.

An important milestone in the development of acoustic event detection and classification has also been some form of feature extraction, that transforms raw waveforms into a sequence of feature vectors that can be used as inputs to deep models~\citep{pmlr-v202-radford23a, gong2021ast, chen2022beats, georgescu2023audiovisual}. Although MFCC spectrograms were demonstrated to work very well for shallow models, modern deep models mainly utilize mel-spectrograms and very recently superlets~\citep{moca2021time}.

\subsection{Downstream Effects of Transforms}

There has been a significant shift from traditional, hand-crafted audio features such as MFCCs to the use of raw audio waveforms and spectrogram representations as inputs for neural networks. ~\citet{wyse2017audio} showed the advantages of spectrogram representations for deep neural networks, particularly their ability to capture both time and frequency information, which is crucial for effectively modeling and generating complex audio signals. Furthermore, multiple works have employed representations based on spectrograms, coupled with convolutional neural networks, and have shown that these work particularly well together~\citep{hershey2017cnn, schmid2023efficient, gong2021psla}.

A class of models is also based on directly processing raw audio~\citep{verma2021audio}. These often involve segmenting the audio input with some window length before converting it into an embedding compatible with the models, rather than producing spectrograms. However, this class of models has seemed to work well primarily for generative applications~\citep{gardner2021mt3}. We do not focus on this class of models for our analysis.

Although there have been studies exploring the classification power of these transforms~\citep{wyse2017audio, ji2020comprehensive, moysis2023music}, none of these works demonstrate the downstream clinical effects of the preprocessing techniques that we focus on in this work. Furthermore, we also analyze additional transforms such as superlets.

\section{Method}

The prospective human patient audio data was collected at a comprehensive stroke center, and the study was approved by the local Research Ethics Board. Subjects were anonymized and the computation occurred locally according to the local institution guidelines.

\begin{figure*}[!t]
    \centering
    \includegraphics{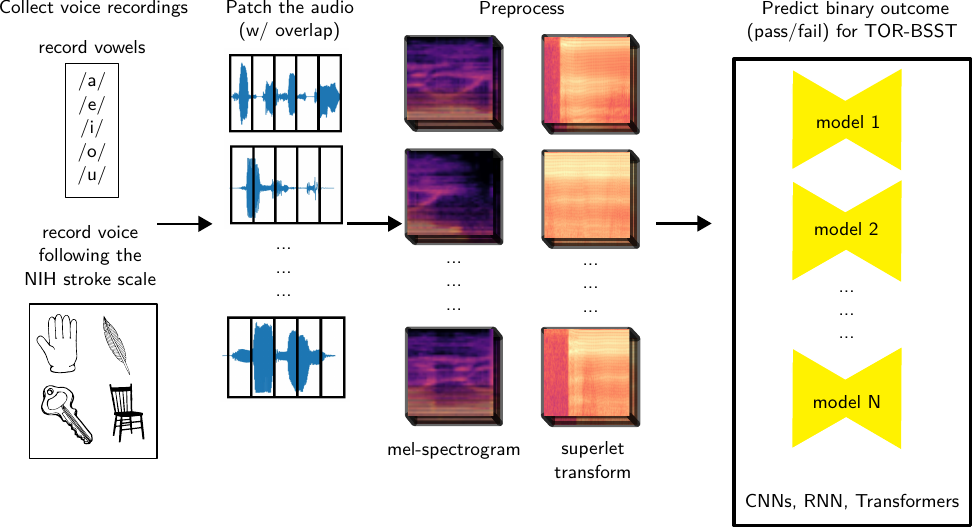}
    \caption{Schematic Overview of the Audio Classification Workflow for Stroke Assessment. The process begins with the collection of voice recordings, including both sustained vowel sounds (/a/, /e/, /i/, /o/, /u/) and speech following the NIHSS, represented by icons for hand, feather, key, chair, etc. These recordings are then segmented using an overlapping patch method to prepare for pre-processing. Subsequently, audio segments are transformed into two types of visual representations: mel-spectrograms and superlet transforms. The final stage involves inputting the processed spectrograms into an array of machine learning models—CNNs, transformers, and RNNs—to predict the outcome of the TOR-BSST$\copyright$ as either ``pass'' or ``fail''.}
    \label{fig:approach}
\end{figure*}

\subsection{Participants}
We enrolled 70 individuals from a comprehensive stroke center, affiliated with the University. These participants were selected during two periods: from June 13, 2022, to January 19, 2023 (epoch 1) and from January 24 to March 4, 2023 (epoch 2), to form training and testing datasets, respectively. Technical problems with the audio recordings resulted in the exclusion of two participants during Epoch 1. Therefore, a total of 68 participant audio samples (with 94\% inter-rater agreement on audio quality by AB and HM) were incorporated into our study. The Toronto Bedside Swallowing Screening Test (TOR-BSST$\copyright$) was administered to all participants as part of standard care, assessing voice changes, repetitive swallows, and dysphonia. Based on this assessment, 27 participants were marked as ``fail'' and 41 as ``pass''. TOR-BSST is a dysphagia screening tool that can be used by operators trained in courses of varying backgrounds~\citet{martino2009toronto}. The study split these participants into two groups: 40 (58.9\%) for training and 28 (41.1\%) for testing. The enrollment was ongoing and based on a randomized approach, targeting admissions to the stroke unit within 72 hours of admission. Every participant gave informed consent, and the research was sanctioned by the stroke center's REB. Inclusion criteria were recent stroke patients proficient in English, able to follow instructions, and without severe aphasia. The excluded included non-English speakers, people with major speech impairments, or medically unstable individuals.

\subsection{Data Collection}
Speech data was divided into two types: a) National Institutes of Health Stroke Scale (NIHSS) speech segments and b) sustained vowel pronunciations. The NIHSS was chosen to avoid bias in test selection as it is commonly used in stroke assessments. NIHSS language tests included continuous speech, sentences, and words. The second dataset comprised vowel sounds (/a/, /e/, /i/, /o/, and /u/), with participants pronouncing each vowel for 3 seconds, three times. This choice was based on evidence showing the uniqueness of vowel sounds in detecting swallowing problems. Data collection was done using an encrypted iPhone 12 and the Voice Recorder app in a real hospital setting. The investigators in charge of data collection, segmentation, and model testing were deliberately kept unaware of each other's activities.

\subsection{Data Analysis}

The initial step involved quality assessment. A three-stage data processing method was adopted, which included segmentation, transformation, and the use of machine learning. Audacity software was used for data segmentation, and a custom Python program transformed segmented audio into Mel-spectrogram image representations (see Figure~\ref{fig:approach}). Given the varied audio durations, a windowing approach ensured consistent Mel-spectrogram image scaling. The Mel spectrograms, renowned for their accuracy in replicating human auditory perception, were critical for our model's success, particularly when compared against evaluations by speech-language professionals. Two distinct Mel-spectrogram images were created for separate machine learning classifier training: RGB and three-channel Mel spectrograms. The latter blends monochrome versions with different FFT lengths. Additionally, Superlet Transform (SLT) was used to create spectrogram images, to assess their performance against mel spectrograms.

\subsection{Pre-training on Public Datasets}
This study delves into the performance enhancement of various networks in a downstream dataset through various pre-training scenarios.

Pre-training on public datasets is crucial for developing machine learning applications tailored to clinical needs, especially when working with small datasets~\citet{rasmy2021med}. This strategy addresses the significant challenge clinicians face in collecting large datasets.

\paragraph{Imagenet}
is a widely recognized dataset crucial for training and evaluating computer vision models. 
\paragraph{AudioSet}
is a comprehensive database featuring 632 categories of audio events in 2,084,320 human-labeled 10-second clips from YouTube. It presents a hierarchical categorization of diverse sounds including those from humans, animals, musical instruments, genres, and everyday life. 

\paragraph{US8K}
is a dataset of 8732 labeled sound excerpts that are up to 4s in length and are taken from~\citep{Salamon:UrbanSound:ACMMM:14}. The sounds are composed of urban sound clips with labels such as air conditioner, car horn, children playing etc.

\paragraph{ESC50}
consists of 2000 audio clips of 5-second-long recordings categorized into 50 classes. These classes fall into the following major categories: animals, natural soundscapes, water sounds, human, non-speech sounds, domestic sounds, urban noises.

\subsection{Exploring Network Architectures for Audio Analysis}


In this section, we explore various neural network architectures, including ConvNeXt and DenseNet for CNN-based models, ConvLSTM2D for temporal data analysis, and Vision Transformer (ViT) and SWIN Transformer for transformer-based models. Additionally, we introduce pre-trained audio feature extractors such as YAMNet, VGGish, and Trill. To address classification tasks, we employ different loss functions and optimizers. For CNN-based models, particularly DenseNet, we implement a hybrid loss function that combines Cross-Entropy and Contrastive Loss. We also incorporate class weights to handle dataset imbalances. Transformer-based models are trained using Cross-Entropy Loss with the inclusion of class weights. The Adam optimizer is chosen for its adaptive learning rate capabilities. Our preprocessing methods involve the use of grayscale audio spectrograms and the conversion of spectrograms into RGB images for select models. Additionally, we explore the use of Superlet transforms in pre-processing. Finally, we evaluate our classifiers using per-participant prediction aggregation (Majority Voting).

\subsubsection{Convolutional Neural Networks (CNNs)}

\paragraph{ConvNeXt} is a recent adaptation of the CNN architecture that has shown impressive results in image classification tasks. Although primarily designed for typical image classification applications, it can be adapted for audio spectrogram analysis, offering potentially more effective feature extraction in sound-based medical diagnostics.

\paragraph{DenseNet} is known for its densely connected convolutional networks, where each layer is connected to every other layer in a feed-forward fashion~\citep{densenet2018}.
The strong performance of the basic DenseNet architecture, as detailed in Table~\ref{tab:detailed_table}, encouraged us to explore its various adaptations, namely:

    \begin{itemize}
        \item DenseNet with binary cross-entropy loss, pretrained on ImageNet (referred to as \textbf{DenseNet}).
        \item DenseNet with a hybrid loss (contrastive loss and binary cross entropy), pre-trained on ImageNet (referred to as \textbf{DenseNet Contrastive}).
        \item DenseNet pre-trained on ImageNet, then on US8K, and applied to our dataset using the hybrid loss (referred to as \textbf{DenseNet Contrastive US8K}).
        \item DenseNet pre-trained on ImageNet and ESC50, later trained on our dataset with the hybrid loss (referred to as \textbf{DenseNet Contrastive ESC50}).
    \end{itemize}

\textit{\textbf{Pre-trained Audio Feature Extractors:}}
\vspace{1mm}

\paragraph{YAMNet} is a CNN that uses the popular MobileNetV1 architecture for the detection of audio events~\citep{howard2017mobilenets}. It is pre-trained on AudioSet to classify various sounds. This pre-training approach streamlines the creation of spectrogram-based filters without requiring extensive proprietary data.

\paragraph{VGGish}~\citep{hershey2017cnn} is a variant of the VGG model~\citep{simonyan2014very}, adapted for audio processing. Initially developed for image classification, the VGG architecture's adaptation to audio allows it to extract meaningful features from sound waves. Like YamNet, it is pre-trained on AudioSet. 

\paragraph{Trill}``TRIpLet Loss network'' is designed for sound event detection and is particularly effective in distinguishing fine-grained acoustic differences. Trill is based on the ResNet50 architecture~\citep{he2015deep} and has shown clear improvements over popular sound classification models~\citep{Shor_2020}.

\subsubsection{Recurrent Neural Networks (RNNs)}
In this paper, we utilized ConvLSTM2D, an amalgamation of CNNs and Long Short-Term Memory (LSTM) networks, designed to capture spatial and temporal relationships in data~\citep{shi2015convolutional}. This is particularly effective for our use case,  which involves generating a series of consecutive spectrograms by splitting a single audio file into multiple fixed-length segments.

\subsubsection{Transformers}

\paragraph{Vision Transformer (ViT)} is tailored for adaptation from large-scale pre-training to fine-tuning on smaller datasets, a process that involves replacing the original MLP head with a new linear layer tailored to the specific class size of the task at hand (2 in our case). This adjustment allows ViTs to be customized for new tasks, without complete retraining~\citep{dosovitskiy2020image}. ViTs utilize self-attention to capture long-range dependencies, a feature that, while powerful, requires extensive training data to achieve the innate perceptual capabilities of CNNs~\citep{zhang2023vitaev2}.

\paragraph{SWIN Transformer} is a variant of ViT, that features a unique ``shifted window'' self-attention mechanism. Unlike ViT, which applies self-attention to the entire sequence of tokens, Swin Transformer performs attention over square-shaped blocks of patches, each block being analogous to a receptive field in convolutional layers. This method enhances hierarchical feature processing with less computational demand. The SWIN Transformer architecture integrates ``merging layers'' for efficient token downsampling and incorporates advanced features such as layer normalization and scaled cosine attention, significantly improving performance and adaptability in transfer learning scenarios~\citep{liu2021swin}.

\textit{\textbf{Pre-trained Audio Feature Extractors:}}
\vspace{1mm}
\input{tables/detailed_comparison}

\paragraph{AST} is a convolution-free, purely attention-based model designed for audio classification~\citep{gong2021ast}. We used the ast-finetuned-audioset-10-10-0.4593 version, pre-trained on the AudioSet dataset. Exploring this model represents a shift from our previous approach, which was more image classification-centric, to one that is specifically designed for audio classification. 

\paragraph{BEATs} model, designed for the extraction of audio features, incorporates 12 transformer encoder layers, 768 hidden states, and 8 attention heads. Pre-trained on AudioSet, BEATs has been evaluated across various audio (AS-2M, AS-20K and ESC-50) and speech (KS1, KS2 and ER) classification tasks, demonstrating its versatility and effectiveness in processing and understanding complex audio data~\citep{chen2022beats}.

\subsection{Loss functions and optimizers}

For our CNN-based models, particularly DenseNet, we implemented a hybrid loss function combining cross-entropy and contrast loss, incorporating class weights to address the imbalance in our data set, notably the under-representation of the ``fail'' class (27 patients) compared to the ``pass'' class (41 patients). Similarly, for Transformer-based models, we applied Cross-Entropy Loss, ensuring class weights were also used. Adam optimizer was chosen for its adaptive learning rate~\citep{kingma2014adam}, optimizing efficiency between models with varied learning rates. This approach was uniformly applied to all trainable parameters to ensure balanced learning dynamics.

\subsection{Preprocessing}
Given the large search space of the many options we wanted to explore in this paper, we chose to approach the problem strategically starting with the most popular approach in the literature: Grayscale (referred to as Mel mono in Table~\ref{tab:options}) audio spectrograms. In order to take advantage of ImageNet pre-training, we ran spectrograms in three different settings, and concatenated the three spectrograms into a single three-channel image. This was then used as input and compared across all models except ConvLSTM2D. The ``Mel mono'' method allowed us to establish a baseline to compare across all feasible models which we then used to refine our training approach and model selection strategy.

For select models, we converted a single-channel spectrogram into an RGB three-channel (``Mel RGB'') image using color-maps and tried this approach on ConvNeXt, Densenet, ViT, Swin Transformer, and ConvLSTM2D network architectures (see Table ~\ref{tab:options}). The strategy here was to make use of the RGB feature extraction abilities of models trained on ImageNet, an RGB-image dataset. 
Additionally,~\citet{palanisamy2020rethinking} discussed a similar approach to convert the spectrogram to an RGB image or choosing different window sizes and hop lengths to create three distinct channels that are concatenated into a single image. They found that on the basis of the baseline model experiments, using mel spectrograms with different window sizes and hop lengths in each channel yielded better performance.

Another preprocessing approach used in this study is Superlet transforms, which is a relatively recent method to transform time-series data into a spectrogram that preserves both time and frequency resolution. Similarly, the grayscale Superlet spectrogram was converted into an RGB image. This approach was only applied to the best performing models, namely ConvNeXt, Densenet, and Swin Transformer.
\subsection{Evaluation of classifiers}
We evaluated each classifier using the aggregation of prediction by participant (majority voting), treating each participant as a single data point, regardless of the number of associated clips. The majority vote across a participant's clips determines their overall prediction. Various performance metrics including F1 Score, Precision, Recall (Sensitivity), and Specificity, both at the validation and test stages were calculated. The Receiver Operating Characteristic (ROC) curve and confusion matrices were also plotted to visually assess performance of models.

\input{tables/options}

\section{Results}
\input{tables/participant_level}
The evaluation of various models on a participant-level classification task reveals performance differences measured by AUC, Sensitivity (ST), and Specificity (SP).

The ConvNeXt model showcased a robust performance across all three metrics in the ``Mel RGB'' category, achieving an AUC of 0.91, ST of 0.78, and SP of 0.89. It maintained this strong performance in the ``Mel mono'' category, with an AUC of 0.86, ST of 0.78, and SP of 0.79, and in the ``Superlet'' category, with slightly lower scores of 0.74 for AUC, 0.68 for ST, and 0.66 for SP.

DenseNet models also performed well, with the standard DenseNet achieving an AUC of 0.89, the highest ST of 0.89, and SP of 0.79 in ``Mel RGB''. In ``Mel mono'', it scored an AUC of 0.88, ST of 0.78, and SP of 0.74, with consistent performance in ``Superlet'' (AUC 0.74, ST 0.67, SP 0.67). Interestingly, the DenseNet Contrastive model excelled in ``Mel mono'' with an AUC of 0.82, ST of 0.86, and SP of 0.78, suggesting its effectiveness in monochrome settings.

In contrast, the ConvLSTM2D model underperformed in the ``mel RGB'' category, with an AUC of only 0.52, although it had a satisfactory ST of 0.78. However, its SP of 0.11 was notably low, indicating a high rate of false positives.

The SWIN Transformer model demonstrated versatility with competitive scores across all categories. It achieved an AUC of 0.801, ST of 0.78, and SP of 0.68 in ``Mel RGB'', and showed improvement in ``Mel mono'' with an AUC of 0.83, ST of 0.6, and SP of 0.83. In the ``Superlet'' category, it scored an AUC of 0.78, ST of 0.67, and SP of 0.79.

The performance of the AST model in ``Mel mono'' was notably effective, with an AUC of 0.83 and ST of 0.89, but its SP of 0.60 suggests a significant trade-off, with a higher tendency for false positives. According to~\cite{gong2021ast}, the AST model does not require as many epochs to train as the CNN-attention hybrid models, which need significantly more epochs. It is worth noting that the AST model required only 6 epochs of training on our dataset to achieve these metrics, which is fewer compared to the CNN-attention hybrid models and other Transformer models explored in this study that needed significantly more epochs to train.

Other models such as YAMNet, VGGish, Trill, and BEATS were assessed only in the ``Mel mono'' category. VGGish showed promising results with an AUC of 0.82, ST of 0.71, and the highest SP of 0.93 among the CNN-based feature extractors, benefiting from its pretraining on AudioSet.

Finally, a detailed comparison of the best performing models (Table~\ref{tab:detailed_table}) in terms of their statistical performance metrics provides a deeper understanding of their predictive capabilities. The ConvNeXt (Mel RGB) model's AUC confidence interval ranged from 0.77 to 1.0, indicating a high degree of certainty in its classification performance, with a commendable F1 score of 0.84, balancing precision and sensitivity. The precision and recall rates of this model, both at 0.78, suggest a harmonious balance between the positive predictive value and the true positive rate.

On the contrary, DenseNet (``Mel RGB'') showed a wider confidence interval in the AUC of 0.73 to 1.0, reflecting more variability in its performance. Despite this, it had the highest sensitivity of 0.89, which shows its strength in identifying true positives. However, the trade-off is evident in its precision of 0.67, which is lower compared to ConvNeXt, leading to an F1 score of 0.81 that, while high, indicates room for improvement in precision.

The SWIN Transformer (``Mel mono'') had a narrower confidence interval for AUC, ranging from 0.5 to 0.9. This range suggests more uncertainty in the model's performance, which is also reflected in the lowest F1 score of 0.57 among the evaluated models. A sensitivity of 0.6 and a specificity of 0.83 show an imbalance, with the model favoring the correct identification of negatives over positives, as also implied by a lower precision rate of 0.55.

AST (``Mel mono'') showed a strong performance with an AUC confidence interval between 0.68 and 0.98 and an F1 score equal to ConvNeXt at 0.84. The model's high sensitivity at 0.89 is on par with DenseNet (``Mel RGB''), but a lower specificity of 0.60 points to a higher false positive rate.

The DenseNet-contrastive US8K model (``Mel mono'') stood out with an AUC confidence interval of 0.76 to 1.0 and perfect scores for both specificity and precision, both at 1.0. This exceptional performance resulted in the highest F1 score of 0.88, indicating a very strong predictive power where the model excelled both in recognizing true positives and in avoiding false positives.



\section{Discussion}

Our study examines the associations between spectrogram preprocessing techniques and the ensuing performance of audio classification models, underscoring an important consideration for clinical applications: the nuanced efficacy of preprocessing approaches has a significant bearing on leveraging transfer learning. Our works suggests that while RGB preprocessing exhibits superior performance in conjunction with ImageNet pre-training, the ``Mel mono'' approach, when pre-trained on expansive public audio datasets, surpasses RGB's effectiveness. This insight is crucial, suggesting that in clinical settings, where data limitations and intrinsic differences are prevalent, adopting a more standardized and contextually tailored approach to preprocessing could significantly enhance the performance of deep learning models. Moreover, the observed variances in model architecture performance, particularly the robustness of transformer-based models versus traditional CNNs in handling limited training epochs, offer a promising avenue for refining audio classification frameworks. This suggests that through strategic selection of preprocessing techniques and models there may be more optimal audio classification strategies that can improve diagnostics in with heightened accuracy and efficiency in clinical environments. This has implications for voice as a biomarker in stroke and other neurologic conditions in addition to other disease states where data limitations may be intrinsic to the health condition including rare diseases.

The DenseNet model, particularly with its Contrastive US8K variant, excelled by leveraging a hybrid loss combining cross-entropy with a supervised contrastive loss, significantly enhancing specificity and yielding the highest F1 score of 0.88 among the evaluated models. The pre-training and fine-tuning process, progressing from general large-scale audio datasets to specialized clinical data, proved crucial in developing effective feature extractors for audio spectrogram classification.

The transformer-based approach, relatively new in the field of audio analysis, demonstrates the potential of adapting architectures originally designed for other domains, like natural language processing, to audio classification. The AST model's  sensitivity score suggests a strong grasp of relevant audio spectrogram features. It showed training efficiency, achieving optimal results with just 6 epochs, contrasting with CNN-based models requiring more epochs for similar performance. This highlights the potential of transformer pre-trained models in audio classification, even with limited training epochs.

We observed variations in performance, partly explained by intrinsic differences and model architecture's ability, within limited data, to classify spectrogram images.

Surprisingly, the RGB preprocessing approach outperformed the grayscale triple channel approach when using the ImageNet pre-training. Theoretically, concatenating three grayscale spectrograms constructed from different mel transform settings should outperform a single mel spectrogram transformed into an RGB image through color mapping. However, our results suggest that the convolutional layers of models pre-trained on ImageNet might be better attuned to the features present in RGB images. This had not been widely discussed in the literature before and we noticed some variations where some papers use grayscale images~\citep{chen2022beats, gong2021ast, howard2017mobilenets, mu2021environmental} , and some would use RGB representations of the spectrograms~\citep{aykanat2017classification, zaman2023survey}. This surprising performance could be attributed to the ImageNet pre-trained models' well-tuned and generalized filters that extract features from structure as well as color. Such generalized filters trained on massive datasets are robust to overfitting and could explain the improved performance on the RGB input.

It is crucial to recognize the impact of confounding variables, especially in a small dataset. Key factors such as age, gender, stroke severity and type, medical comorbidities, medications, cognitive function, psychological factors, rehabilitation history, time since stroke, environmental factors, and nutritional status have an effect on patient performance. Stratifying different populations based on these factors would be an important next step.

\section{Conclusion}

In summary, our study underscores the effectiveness of modern CNN architectures, such as DenseNet and ConvNeXt, in the field of clinical audio classification. These architectures demonstrate robustness, often rivaling or even surpassing the capabilities of transformer models, particularly in scenarios involving small datasets. A key factor in this success is the strategic use of open-source pre-trained weights, which not only accelerates the development process but also significantly enhances model accuracy.

A cornerstone of our research involved the prospective collection of two first-of-their-kind patient audio datasets from stroke patients. We introduce the use of an NIHSS-based audio dataset, a novel collection that captures continuous speech, sentences, and words based on the NIHSS. This well-established test for the assessment of stroke in emergency departments provides invaluable data to develop audio classification models tailored to clinical needs. Additionally, we presented the Vowel dataset, a unique compilation of sustained vowel sounds from patients, offering new insights into the analysis of swallowing disorders.

Leveraging open datasets for pre-training enables generalized feature learning, essential for subsequent fine-tuning on specific datasets. Our study highlights the effectiveness of a multistage training and fine-tuning process for gradual model adaptation and improved performance. The influence of preprocessing techniques, such as Mel RGB, Mel mono, and Superlet, on model performance is significant and requires careful selection. Temporal segregation between training and testing datasets is crucial to prevent data leakage and enhance model generalization. Additionally, our findings underscore the potential of audio as a robust predictor of clinically relevant information, exemplified by our successful prediction of swallowing status based solely on audio data, suggesting promising applications in clinical settings.

Through nuanced consideration in data handling and model training, our research contributes to the advancement of clinical audio classification, with promising implications for its application in various conditions both in neurologic conditions and beyond.

\acks{We thank anonymous reviewers of the CHIL Conference for their insightful suggestions which we incorporate in this work.}

\bibliography{chil-sample}

\end{document}

%% file: math_commands.tex
\def\eqref#1{equation~\ref{#1}}

\def\1{\bm{1}}

\DeclareMathAlphabet{\mathsfit}{\encodingdefault}{\sfdefault}{m}{sl}
\SetMathAlphabet{\mathsfit}{bold}{\encodingdefault}{\sfdefault}{bx}{n}



%% file: tables/detailed_comparison.tex
\begin{table*}[!t]
\caption{Detailed comparison of model performance on participant level classification task (AUC=Area Under the ROC Curve. ST = Sensitivity. SP = Specificity).}
\label{tab:detailed_table}
\centering
\begin{tabular}{@{}lccccc@{}}
\toprule
Method                     & AUC 95\% CI      & ST (Recall) & SP   & Precision & F1 score \\ \midrule
ConvNeXT(RGB)                   & 0.77 - 1.0  & 0.78        & 0.89 & 0.78      & 0.84     \\
DenseNet(RGB)                   & 0.73 - 1.0  & 0.89        & 0.79 & 0.67      & 0.81     \\
SWIN Transformer(mono)           & 0.5-0.9     & 0.6         & 0.83 & 0.55      & 0.57     \\
AST(mono)                        & 0.68 - 0.98 & 0.89        & 0.60 & 0.80      & 0.84     \\
DenseNet-Constrastive US8K(mono) & 0.76-1.0    & 0.78        & 1.0  & 1.0       & 0.88     \\ \bottomrule
\end{tabular}
\end{table*}

%% file: tables/options.tex
\begin{table}
\caption{List of models evaluated for performance comparison alongside the different pre-processing options used for each.}
\label{tab:options}
\begin{tabularx}{\linewidth}{Xccc}
\toprule
Method                     & Mel RGB & Mel mono & Superlet \\ \midrule
YAMNet                     &         & \checkmark        &          \\
VGGish                     &         & \checkmark        &          \\
Trill                      &         & \checkmark        &          \\
BEATS                      &         & \checkmark        &          \\
ConvNeXT                   & \checkmark       & \checkmark        & \checkmark        \\
DenseNet                   & \checkmark       & \checkmark        & \checkmark        \\
DenseNet-Contrastive       &         & \checkmark        &          \\
DenseNet-Contrastive US8K  &         & \checkmark        &          \\
DenseNet-Contrastive ESC50 &         & \checkmark        &          \\
ConvLSTM2D                 & \checkmark       &          &          \\
ViT                        & \checkmark       & \checkmark        &          \\
SWIN             & \checkmark       & \checkmark        & \checkmark        \\
AST                        &         & \checkmark        &          \\ \bottomrule
\end{tabularx}
\end{table} 

%% file: tables/participant_level.tex
\begin{table*}[]
\caption{Overview of model performance on participant level classification task (where AUC represents Area Under the ROC Curve, ST represents the Sensitivity, and SP represents the Specificity).}
\label{tab:participant level}
\centering
\begin{tabular}{lrrrrrrrrr}
\toprule
Method                      & \multicolumn{3}{c}{Mel RGB} & \multicolumn{3}{c}{Mel mono} & \multicolumn{3}{c}{Superlet} \\ \midrule
                            & AUC     & ST     & SP    & AUC    & ST     & SP     & AUC    & ST     & SP     \\ \cline{2-10} 
YAMNet                      &  -  &  -  &  -  &                      0.69 &                      0.71 &                      0.79 &  -  &  -  &  -  \\
VGGish                      &  -  &  -  &  -  &                      0.82 &                      0.71 &                      0.93 &  -  &  -  &  -  \\
Trill                       &  -  &  -  &  -  &                      0.71 &                      0.57 &                      0.86 &  -  &  -  &  -  \\
BEATS                       &  -  &  -  &  -  &                      0.44 &                      0.57 &                      0.55 &  -  &  -  &  -  \\
ConvNeXt                    & \cellcolor{tabsecond}0.91 &  \cellcolor{tabthird}0.78 & \cellcolor{tabsecond}0.89 &                      0.86 &                      0.78 &                      0.79 &  \cellcolor{tabthird}0.74 & \cellcolor{tabsecond}0.68 &                      0.66 \\
DenseNet                    &  \cellcolor{tabthird}0.89 & \cellcolor{tabsecond}0.89 &  \cellcolor{tabthird}0.79 &  \cellcolor{tabthird}0.88 &                      0.78 &                      0.74 &  \cellcolor{tabthird}0.74 &  \cellcolor{tabthird}0.67 &  \cellcolor{tabthird}0.67 \\
DenseNet Contrastive        &  -  &  -  &  -  &                      0.82 &  \cellcolor{tabthird}0.86 &                      0.78 &  -  &  -  &  -  \\
DenseNet Contrastive US8K   &  -  &  -  &  -  & \cellcolor{tabsecond}0.89 &                      0.78 & \cellcolor{tabsecond}1.00 &  -  &  -  &  -  \\
DenseNet Constrastive ESC50 &  -  &  -  &  -  &                      0.75 &                      0.71 &                      0.78 &  -  &  -  &  -  \\
ConvLSTM2D                  &                      0.52 &  \cellcolor{tabthird}0.78 &                      0.11 &  -  &  -  &  -  &  -  &  -  &  -  \\
ViT                         &                      0.79 &                      0.67 & \cellcolor{tabsecond}0.89 &                      0.84 &                      0.40 &  \cellcolor{tabthird}0.94 &  -  &  -  &  -  \\
SWIN Transformer            &                      0.80 &  \cellcolor{tabthird}0.78 &                      0.68 &                      0.83 &                      0.60 &                      0.83 & \cellcolor{tabsecond}0.78 &  \cellcolor{tabthird}0.67 & \cellcolor{tabsecond}0.79 \\
AST                         &  -  &  -  &  -  &                      0.83 & \cellcolor{tabsecond}0.89 &                      0.60 &  -  &  -  &  - \\
\bottomrule  
\end{tabular}
\end{table*}

%% file: chil-manuscript.bbl
\begin{thebibliography}{69}
\providecommand{\natexlab}[1]{#1}
\providecommand{\url}[1]{\texttt{#1}}
\expandafter\ifx\csname urlstyle\endcsname\relax
  \providecommand{\doi}[1]{doi: #1}\else
  \providecommand{\doi}{doi: \begingroup \urlstyle{rm}\Url}\fi

\bibitem[Alqudaihi et~al.(2021)Alqudaihi, Aslam, Khan, Almuhaideb, Alsunaidi, Ibrahim, Alhaidari, Shaikh, Alsenbel, Alalharith, Alharthi, Alghamdi, and Alshahrani]{9486929}
Kawther~S. Alqudaihi, Nida Aslam, Irfan~Ullah Khan, Abdullah~M. Almuhaideb, Shikah~J. Alsunaidi, Nehad M. Abdel~Rahman Ibrahim, Fahd~A. Alhaidari, Fatema~S. Shaikh, Yasmine~M. Alsenbel, Dima~M. Alalharith, Hajar~M. Alharthi, Wejdan~M. Alghamdi, and Mohammed~S. Alshahrani.
\newblock Cough sound detection and diagnosis using artificial intelligence techniques: Challenges and opportunities.
\newblock \emph{IEEE Access}, 9:\penalty0 102327--102344, 2021.
\newblock \doi{10.1109/ACCESS.2021.3097559}.

\bibitem[Aykanat et~al.(2017)Aykanat, K{\i}l{\i}{\c{c}}, Kurt, and Saryal]{aykanat2017classification}
Murat Aykanat, {\"O}zkan K{\i}l{\i}{\c{c}}, Bahar Kurt, and Sevgi Saryal.
\newblock Classification of lung sounds using convolutional neural networks.
\newblock \emph{EURASIP Journal on Image and Video Processing}, 2017\penalty0 (1):\penalty0 1--9, 2017.

\bibitem[Baevski et~al.(2020)Baevski, Zhou, Mohamed, and Auli]{NEURIPS2020_92d1e1eb}
Alexei Baevski, Yuhao Zhou, Abdelrahman Mohamed, and Michael Auli.
\newblock wav2vec 2.0: A framework for self-supervised learning of speech representations.
\newblock In H.~Larochelle, M.~Ranzato, R.~Hadsell, M.F. Balcan, and H.~Lin, editors, \emph{Advances in Neural Information Processing Systems}, volume~33, pages 12449--12460. Curran Associates, Inc., 2020.
\newblock URL \url{https://proceedings.neurips.cc/paper_files/paper/2020/file/92d1e1eb1cd6f9fba3227870bb6d7f07-Paper.pdf}.

\bibitem[B{\"o}ck and Schedl(2011)]{bock2011enhanced}
Sebastian B{\"o}ck and Markus Schedl.
\newblock Enhanced beat tracking with context-aware neural networks.
\newblock In \emph{Proc. Int. Conf. Digital Audio Effects}, pages 135--139, 2011.

\bibitem[Chen et~al.(2022)Chen, Wu, Wang, Liu, Tompkins, Chen, and Wei]{chen2022beats}
Sanyuan Chen, Yu~Wu, Chengyi Wang, Shujie Liu, Daniel Tompkins, Zhuo Chen, and Furu Wei.
\newblock Beats: Audio pre-training with acoustic tokenizers.
\newblock \emph{arXiv preprint arXiv:2212.09058}, 2022.

\bibitem[Choi et~al.(2018)Choi, Fazekas, Sandler, and Cho]{8553106}
Keunwoo Choi, György Fazekas, Mark Sandler, and Kyunghyun Cho.
\newblock A comparison of audio signal preprocessing methods for deep neural networks on music tagging.
\newblock In \emph{2018 26th European Signal Processing Conference (EUSIPCO)}, pages 1870--1874, 2018.
\newblock \doi{10.23919/EUSIPCO.2018.8553106}.

\bibitem[Deng et~al.(2009)Deng, Dong, Socher, Li, Li, and Fei-Fei]{deng2009imagenet}
Jia Deng, Wei Dong, Richard Socher, Li-Jia Li, Kai Li, and Li~Fei-Fei.
\newblock Imagenet: A large-scale hierarchical image database.
\newblock In \emph{2009 IEEE conference on computer vision and pattern recognition}, pages 248--255. Ieee, 2009.

\bibitem[Dhanalakshmi et~al.(2009)Dhanalakshmi, Palanivel, and Ramalingam]{dhanalakshmi2009classification}
P~Dhanalakshmi, S~Palanivel, and Vennila Ramalingam.
\newblock Classification of audio signals using svm and rbfnn.
\newblock \emph{Expert systems with applications}, 36\penalty0 (3):\penalty0 6069--6075, 2009.

\bibitem[Dosovitskiy et~al.(2020)Dosovitskiy, Beyer, Kolesnikov, Weissenborn, Zhai, Unterthiner, Dehghani, Minderer, Heigold, Gelly, et~al.]{dosovitskiy2020image}
Alexey Dosovitskiy, Lucas Beyer, Alexander Kolesnikov, Dirk Weissenborn, Xiaohua Zhai, Thomas Unterthiner, Mostafa Dehghani, Matthias Minderer, Georg Heigold, Sylvain Gelly, et~al.
\newblock An image is worth 16x16 words: Transformers for image recognition at scale.
\newblock \emph{arXiv preprint arXiv:2010.11929}, 2020.

\bibitem[Dwivedi et~al.(2018)Dwivedi, Imtiaz, and Rodriguez-Villegas]{dwivedi2018algorithms}
Amit~Krishna Dwivedi, Syed~Anas Imtiaz, and Esther Rodriguez-Villegas.
\newblock Algorithms for automatic analysis and classification of heart sounds--a systematic review.
\newblock \emph{IEEE Access}, 7:\penalty0 8316--8345, 2018.

\bibitem[Emmanuel(2012)]{emmanuel2012review}
Babatunde~S Emmanuel.
\newblock A review of signal processing techniques for heart sound analysis in clinical diagnosis.
\newblock \emph{Journal of medical engineering \& technology}, 36\penalty0 (6):\penalty0 303--307, 2012.

\bibitem[Fagherazzi et~al.(2021)Fagherazzi, Fischer, Ismael, and Despotovic]{fagherazzi2021voice}
Guy Fagherazzi, Aur{\'e}lie Fischer, Muhannad Ismael, and Vladimir Despotovic.
\newblock Voice for health: the use of vocal biomarkers from research to clinical practice.
\newblock \emph{Digital biomarkers}, 5\penalty0 (1):\penalty0 78--88, 2021.

\bibitem[García-Ordás et~al.(2020)García-Ordás, Benítez-Andrades, García-Rodríguez, Benavides, and Alaiz-Moretón]{s20041214}
María~Teresa García-Ordás, José~Alberto Benítez-Andrades, Isaías García-Rodríguez, Carmen Benavides, and Héctor Alaiz-Moretón.
\newblock Detecting respiratory pathologies using convolutional neural networks and variational autoencoders for unbalancing data.
\newblock \emph{Sensors}, 20\penalty0 (4), 2020.
\newblock ISSN 1424-8220.
\newblock \doi{10.3390/s20041214}.
\newblock URL \url{https://www.mdpi.com/1424-8220/20/4/1214}.

\bibitem[Gardner et~al.(2021)Gardner, Simon, Manilow, Hawthorne, and Engel]{gardner2021mt3}
Josh Gardner, Ian Simon, Ethan Manilow, Curtis Hawthorne, and Jesse Engel.
\newblock Mt3: Multi-task multitrack music transcription.
\newblock \emph{arXiv preprint arXiv:2111.03017}, 2021.

\bibitem[Gavriely et~al.(1994)Gavriely, Nissan, Cugell, and Rubin]{gavriely1994respiratory}
N~Gavriely, M~Nissan, DW~Cugell, and AHE Rubin.
\newblock Respiratory health screening using pulmonary function tests and lung sound analysis.
\newblock \emph{European Respiratory Journal}, 7\penalty0 (1):\penalty0 35--42, 1994.

\bibitem[Gemmeke et~al.(2017)Gemmeke, Ellis, Freedman, Jansen, Lawrence, Moore, Plakal, and Ritter]{gemmeke2017audio}
Jort~F Gemmeke, Daniel~PW Ellis, Dylan Freedman, Aren Jansen, Wade Lawrence, R~Channing Moore, Manoj Plakal, and Marvin Ritter.
\newblock Audio set: An ontology and human-labeled dataset for audio events.
\newblock In \emph{2017 IEEE international conference on acoustics, speech and signal processing (ICASSP)}, pages 776--780. IEEE, 2017.

\bibitem[Georgescu et~al.(2023)Georgescu, Fonseca, Ionescu, Lucic, Schmid, and Arnab]{georgescu2023audiovisual}
Mariana-Iuliana Georgescu, Eduardo Fonseca, Radu~Tudor Ionescu, Mario Lucic, Cordelia Schmid, and Anurag Arnab.
\newblock Audiovisual masked autoencoders.
\newblock In \emph{Proceedings of the IEEE/CVF International Conference on Computer Vision}, pages 16144--16154, 2023.

\bibitem[Gimeno et~al.(2020)Gimeno, Vi{\~n}als, Ortega, Miguel, and Lleida]{gimeno2020multiclass}
Pablo Gimeno, Ignacio Vi{\~n}als, Alfonso Ortega, Antonio Miguel, and Eduardo Lleida.
\newblock Multiclass audio segmentation based on recurrent neural networks for broadcast domain data.
\newblock \emph{EURASIP Journal on Audio, Speech, and Music Processing}, 2020:\penalty0 1--19, 2020.

\bibitem[Gong et~al.(2021{\natexlab{a}})Gong, Chung, and Glass]{gong2021ast}
Yuan Gong, Yu-An Chung, and James Glass.
\newblock Ast: Audio spectrogram transformer, 2021{\natexlab{a}}.

\bibitem[Gong et~al.(2021{\natexlab{b}})Gong, Chung, and Glass]{gong2021psla}
Yuan Gong, Yu-An Chung, and James Glass.
\newblock Psla: Improving audio tagging with pretraining, sampling, labeling, and aggregation.
\newblock \emph{IEEE/ACM Transactions on Audio, Speech, and Language Processing}, 29:\penalty0 3292--3306, 2021{\natexlab{b}}.

\bibitem[Gong et~al.(2022)Gong, Rouditchenko, Liu, Harwath, Karlinsky, Kuehne, and Glass]{gong2022contrastive}
Yuan Gong, Andrew Rouditchenko, Alexander~H Liu, David Harwath, Leonid Karlinsky, Hilde Kuehne, and James Glass.
\newblock Contrastive audio-visual masked autoencoder.
\newblock \emph{arXiv preprint arXiv:2210.07839}, 2022.

\bibitem[He et~al.(2015)He, Zhang, Ren, and Sun]{he2015deep}
Kaiming He, Xiangyu Zhang, Shaoqing Ren, and Jian Sun.
\newblock Deep residual learning for image recognition. corr abs/1512.03385 (2015), 2015.

\bibitem[Hermansky et~al.(2000)Hermansky, Ellis, and Sharma]{hermansky2000tandem}
Hynek Hermansky, Daniel~PW Ellis, and Sangita Sharma.
\newblock Tandem connectionist feature extraction for conventional hmm systems.
\newblock In \emph{2000 IEEE international conference on acoustics, speech, and signal processing. Proceedings (Cat. No. 00CH37100)}, volume~3, pages 1635--1638. IEEE, 2000.

\bibitem[Hershey et~al.(2017)Hershey, Chaudhuri, Ellis, Gemmeke, Jansen, Moore, Plakal, Platt, Saurous, Seybold, et~al.]{hershey2017cnn}
Shawn Hershey, Sourish Chaudhuri, Daniel~PW Ellis, Jort~F Gemmeke, Aren Jansen, R~Channing Moore, Manoj Plakal, Devin Platt, Rif~A Saurous, Bryan Seybold, et~al.
\newblock Cnn architectures for large-scale audio classification.
\newblock In \emph{2017 ieee international conference on acoustics, speech and signal processing (icassp)}, pages 131--135. IEEE, 2017.

\bibitem[Holzapfel and Stylianou(2008)]{holzapfel2008musical}
Andre Holzapfel and Yannis Stylianou.
\newblock Musical genre classification using nonnegative matrix factorization-based features.
\newblock \emph{IEEE Transactions on Audio, Speech, and Language Processing}, 16\penalty0 (2):\penalty0 424--434, 2008.

\bibitem[Howard et~al.(2017)Howard, Zhu, Chen, Kalenichenko, Wang, Weyand, Andreetto, and Adam]{howard2017mobilenets}
Andrew~G Howard, Menglong Zhu, Bo~Chen, Dmitry Kalenichenko, Weijun Wang, Tobias Weyand, Marco Andreetto, and Hartwig Adam.
\newblock Mobilenets: Efficient convolutional neural networks for mobile vision applications.
\newblock \emph{arXiv preprint arXiv:1704.04861}, 2017.

\bibitem[Huang et~al.(2018)Huang, Liu, Laurens~van der, and Q.~Weinberger]{densenet2018}
Gao Huang, Zhuang Liu, Maaten Laurens~van der, and Kilian Q.~Weinberger.
\newblock Densely connected convolutional networks.
\newblock \emph{arXiv preprint arXiv:1608.06993}, 2018.
\newblock URL \url{https://arxiv.org/pdf/1608.06993.pdf}.

\bibitem[Huang et~al.(2022)Huang, Xu, Li, Baevski, Auli, Galuba, Metze, and Feichtenhofer]{huang2022masked}
Po-Yao Huang, Hu~Xu, Juncheng Li, Alexei Baevski, Michael Auli, Wojciech Galuba, Florian Metze, and Christoph Feichtenhofer.
\newblock Masked autoencoders that listen.
\newblock \emph{arXiv preprint arXiv:2207.06405}, 2022.

\bibitem[Jaegle et~al.(2021)Jaegle, Gimeno, Brock, Vinyals, Zisserman, and Carreira]{jaegle2021perceiver}
Andrew Jaegle, Felix Gimeno, Andy Brock, Oriol Vinyals, Andrew Zisserman, and Joao Carreira.
\newblock Perceiver: General perception with iterative attention.
\newblock In \emph{International conference on machine learning}, pages 4651--4664. PMLR, 2021.

\bibitem[Ji et~al.(2020)Ji, Luo, and Yang]{ji2020comprehensive}
Shulei Ji, Jing Luo, and Xinyu Yang.
\newblock A comprehensive survey on deep music generation: Multi-level representations, algorithms, evaluations, and future directions.
\newblock \emph{arXiv preprint arXiv:2011.06801}, 2020.

\bibitem[Juang(1984)]{6770768}
B.-H. Juang.
\newblock On the hidden markov model and dynamic time warping for speech recognition — a unified view.
\newblock \emph{AT\&T Bell Laboratories Technical Journal}, 63\penalty0 (7):\penalty0 1213--1243, 1984.
\newblock \doi{10.1002/j.1538-7305.1984.tb00034.x}.

\bibitem[Kingma and Ba(2014)]{kingma2014adam}
Diederik~P Kingma and Jimmy Ba.
\newblock Adam: A method for stochastic optimization.
\newblock \emph{arXiv preprint arXiv:1412.6980}, 2014.
\newblock URL \url{https://arxiv.org/abs/1412.6980}.

\bibitem[Kong et~al.(2020)Kong, Ping, Huang, Zhao, and Catanzaro]{kong2020diffwave}
Zhifeng Kong, Wei Ping, Jiaji Huang, Kexin Zhao, and Bryan Catanzaro.
\newblock Diffwave: A versatile diffusion model for audio synthesis.
\newblock \emph{arXiv preprint arXiv:2009.09761}, 2020.

\bibitem[Koutini et~al.(2021)Koutini, Schl{\"u}ter, Eghbal-Zadeh, and Widmer]{koutini2021efficient}
Khaled Koutini, Jan Schl{\"u}ter, Hamid Eghbal-Zadeh, and Gerhard Widmer.
\newblock Efficient training of audio transformers with patchout.
\newblock \emph{arXiv preprint arXiv:2110.05069}, 2021.

\bibitem[Kwah and Diong(2014)]{kwah2014national}
Li~Khim Kwah and Joanna Diong.
\newblock National institutes of health stroke scale (nihss).
\newblock \emph{Journal of physiotherapy}, 2014.

\bibitem[Larson et~al.(2012)Larson, Comina, Gilman, Tracey, Bravard, and L{\'o}pez]{larson2012validation}
Sandra Larson, Germ{\'a}n Comina, Robert~H Gilman, Brian~H Tracey, Marjory Bravard, and Jos{\'e}~W L{\'o}pez.
\newblock Validation of an automated cough detection algorithm for tracking recovery of pulmonary tuberculosis patients.
\newblock 2012.

\bibitem[Lee and Han(2021)]{lee2021nu}
Junhyeok Lee and Seungu Han.
\newblock Nu-wave: A diffusion probabilistic model for neural audio upsampling.
\newblock \emph{arXiv preprint arXiv:2104.02321}, 2021.

\bibitem[Liu et~al.(2021)Liu, Lin, Cao, Hu, Wei, Zhang, Lin, and Guo]{liu2021swin}
Ze~Liu, Yutong Lin, Yue Cao, Han Hu, Yixuan Wei, Zheng Zhang, Stephen Lin, and Baining Guo.
\newblock Swin transformer: Hierarchical vision transformer using shifted windows.
\newblock In \emph{Proceedings of the IEEE/CVF international conference on computer vision}, pages 10012--10022, 2021.

\bibitem[Martino et~al.(2009)Martino, Silver, Teasell, Bayley, Nicholson, Streiner, and Diamant]{martino2009toronto}
Rosemary Martino, Frank Silver, Robert Teasell, Mark Bayley, Gordon Nicholson, David~L Streiner, and Nicholas~E Diamant.
\newblock The toronto bedside swallowing screening test (tor-bsst) development and validation of a dysphagia screening tool for patients with stroke.
\newblock \emph{Stroke}, 40\penalty0 (2):\penalty0 555--561, 2009.

\bibitem[Moca et~al.(2021)Moca, B{\^a}rzan, Nagy-D{\u{a}}b{\^a}can, and Mureșan]{moca2021time}
Vasile~V Moca, Harald B{\^a}rzan, Adriana Nagy-D{\u{a}}b{\^a}can, and Raul~C Mureșan.
\newblock Time-frequency super-resolution with superlets.
\newblock \emph{Nature communications}, 12\penalty0 (1):\penalty0 337, 2021.

\bibitem[Moedomo et~al.(2012)Moedomo, Mardiyanto, Ahmad, Alisjahbana, and Djatmiko]{moedomo2012breath}
Ria~Lestari Moedomo, M~Sukrisno Mardiyanto, Munawar Ahmad, Bachti Alisjahbana, and Tjahjono Djatmiko.
\newblock The breath sound analysis for diseases diagnosis and stress measurement.
\newblock In \emph{2012 International Conference on System Engineering and Technology (ICSET)}, pages 1--6. IEEE, 2012.

\bibitem[Moysis et~al.(2023)Moysis, Iliadis, Sotiroudis, Boursianis, Papadopoulou, Kokkinidis, Volos, Sarigiannidis, Nikolaidis, and Goudos]{moysis2023music}
Lazaros Moysis, Lazaros~Alexios Iliadis, Sotirios~P Sotiroudis, Achilles~D Boursianis, Maria~S Papadopoulou, Konstantinos-Iraklis~D Kokkinidis, Christos Volos, Panagiotis Sarigiannidis, Spiridon Nikolaidis, and Sotirios~K Goudos.
\newblock Music deep learning: Deep learning methods for music signal processing-a review of the state-of-the-art.
\newblock \emph{IEEE Access}, 2023.

\bibitem[Mu et~al.(2021)Mu, Yin, Huang, Xu, and Du]{mu2021environmental}
Wenjie Mu, Bo~Yin, Xianqing Huang, Jiali Xu, and Zehua Du.
\newblock Environmental sound classification using temporal-frequency attention based convolutional neural network.
\newblock \emph{Scientific Reports}, 11\penalty0 (1):\penalty0 21552, 2021.

\bibitem[Nilsson et~al.(2002)Nilsson, Gustaftson, Andersen, and Kleijn]{nilsson2002gaussian}
Mattias Nilsson, Harald Gustaftson, S{\o}ren~Vang Andersen, and W~Bastiaan Kleijn.
\newblock Gaussian mixture model based mutual information estimation between frequency bands in speech.
\newblock In \emph{2002 IEEE International Conference on Acoustics, Speech, and Signal Processing}, volume~1, pages I--525. IEEE, 2002.

\bibitem[Ozerov and F{\'e}votte(2009)]{ozerov2009multichannel}
Alexey Ozerov and C{\'e}dric F{\'e}votte.
\newblock Multichannel nonnegative matrix factorization in convolutive mixtures for audio source separation.
\newblock \emph{IEEE transactions on audio, speech, and language processing}, 18\penalty0 (3):\penalty0 550--563, 2009.

\bibitem[Pahar et~al.(2021)Pahar, Klopper, Reeve, Warren, Theron, and Niesler]{Pahar_2021}
Madhurananda Pahar, Marisa Klopper, Byron Reeve, Rob Warren, Grant Theron, and Thomas Niesler.
\newblock Automatic cough classification for tuberculosis screening in a real-world environment.
\newblock \emph{Physiological Measurement}, 42\penalty0 (10):\penalty0 105014, nov 2021.
\newblock \doi{10.1088/1361-6579/ac2fb8}.
\newblock URL \url{https://dx.doi.org/10.1088/1361-6579/ac2fb8}.

\bibitem[Palanisamy et~al.(2020)Palanisamy, Singhania, and Yao]{palanisamy2020rethinking}
Kamalesh Palanisamy, Dipika Singhania, and Angela Yao.
\newblock Rethinking cnn models for audio classification.
\newblock \emph{arXiv preprint arXiv:2007.11154}, 2020.

\bibitem[Phan et~al.(2017)Phan, Koch, Katzberg, Maass, Mazur, and Mertins]{phan2017audio}
Huy Phan, Philipp Koch, Fabrice Katzberg, Marco Maass, Radoslaw Mazur, and Alfred Mertins.
\newblock Audio scene classification with deep recurrent neural networks.
\newblock \emph{arXiv preprint arXiv:1703.04770}, 2017.

\bibitem[Piczak()]{piczak2015dataset}
Karol~J. Piczak.
\newblock {ESC}: {Dataset} for {Environmental Sound Classification}.
\newblock In \emph{Proceedings of the 23rd {Annual ACM Conference} on {Multimedia}}, pages 1015--1018. {ACM Press}.
\newblock ISBN 978-1-4503-3459-4.
\newblock \doi{10.1145/2733373.2806390}.
\newblock URL \url{http://dl.acm.org/citation.cfm?doid=2733373.2806390}.

\bibitem[Radford et~al.(2023)Radford, Kim, Xu, Brockman, Mcleavey, and Sutskever]{pmlr-v202-radford23a}
Alec Radford, Jong~Wook Kim, Tao Xu, Greg Brockman, Christine Mcleavey, and Ilya Sutskever.
\newblock Robust speech recognition via large-scale weak supervision.
\newblock In Andreas Krause, Emma Brunskill, Kyunghyun Cho, Barbara Engelhardt, Sivan Sabato, and Jonathan Scarlett, editors, \emph{Proceedings of the 40th International Conference on Machine Learning}, volume 202 of \emph{Proceedings of Machine Learning Research}, pages 28492--28518. PMLR, 23--29 Jul 2023.
\newblock URL \url{https://proceedings.mlr.press/v202/radford23a.html}.

\bibitem[Raphael(1999)]{raphael1999automatic}
Christopher Raphael.
\newblock Automatic segmentation of acoustic musical signals using hidden markov models.
\newblock \emph{IEEE transactions on pattern analysis and machine intelligence}, 21\penalty0 (4):\penalty0 360--370, 1999.

\bibitem[Rasmy et~al.(2021)Rasmy, Xiang, Xie, Tao, and Zhi]{rasmy2021med}
Laila Rasmy, Yang Xiang, Ziqian Xie, Cui Tao, and Degui Zhi.
\newblock Med-bert: pretrained contextualized embeddings on large-scale structured electronic health records for disease prediction.
\newblock \emph{NPJ digital medicine}, 4\penalty0 (1):\penalty0 86, 2021.

\bibitem[Saab et~al.(2023)Saab, Balachandar, Mahdi, and Nashnoush]{masa2023paper}
Rami Saab, Arjun Balachandar, Hamza Mahdi, and Eptehal Nashnoush.
\newblock Machine-learning assisted swallowing assessment: a deep learning-based quality improvement tool to screen for post-stroke dysphagia.
\newblock \emph{Frontiers in Neuroscience}, 17, 2023.
\newblock \doi{10.3389/fnins.2023.1302132}.
\newblock URL \url{https://www.frontiersin.org/articles/10.3389/fnins.2023.1302132/full}.

\bibitem[Sakoe and Chiba(1978)]{1163055}
H.~Sakoe and S.~Chiba.
\newblock Dynamic programming algorithm optimization for spoken word recognition.
\newblock \emph{IEEE Transactions on Acoustics, Speech, and Signal Processing}, 26\penalty0 (1):\penalty0 43--49, 1978.
\newblock \doi{10.1109/TASSP.1978.1163055}.

\bibitem[Salamon et~al.(2014)Salamon, Jacoby, and Bello]{Salamon:UrbanSound:ACMMM:14}
J.~Salamon, C.~Jacoby, and J.~P. Bello.
\newblock A dataset and taxonomy for urban sound research.
\newblock In \emph{22nd {ACM} International Conference on Multimedia (ACM-MM'14)}, pages 1041--1044, Orlando, FL, USA, Nov. 2014.

\bibitem[Salvador and Chan(2007)]{salvador2007toward}
Stan Salvador and Philip Chan.
\newblock Toward accurate dynamic time warping in linear time and space.
\newblock \emph{Intelligent Data Analysis}, 11\penalty0 (5):\penalty0 561--580, 2007.

\bibitem[Schmid et~al.(2023)Schmid, Koutini, and Widmer]{schmid2023efficient}
Florian Schmid, Khaled Koutini, and Gerhard Widmer.
\newblock Efficient large-scale audio tagging via transformer-to-cnn knowledge distillation.
\newblock In \emph{ICASSP 2023-2023 IEEE International Conference on Acoustics, Speech and Signal Processing (ICASSP)}, pages 1--5. IEEE, 2023.

\bibitem[Seide et~al.(2011)Seide, Li, Chen, and Yu]{6163899}
Frank Seide, Gang Li, Xie Chen, and Dong Yu.
\newblock Feature engineering in context-dependent deep neural networks for conversational speech transcription.
\newblock In \emph{2011 IEEE Workshop on Automatic Speech Recognition \& Understanding}, pages 24--29, 2011.
\newblock \doi{10.1109/ASRU.2011.6163899}.

\bibitem[Sello et~al.(2008)Sello, Strambi, De~Michele, and Ambrosino]{sello2008respiratory}
Stefano Sello, Soo-kyung Strambi, Gennaro De~Michele, and Nicolino Ambrosino.
\newblock Respiratory sound analysis in healthy and pathological subjects: A wavelet approach.
\newblock \emph{Biomedical Signal Processing and Control}, 3\penalty0 (3):\penalty0 181--191, 2008.

\bibitem[Shi et~al.(2015)Shi, Chen, Wang, Yeung, Wong, and Woo]{shi2015convolutional}
Xingjian Shi, Zhourong Chen, Hao Wang, DY~Yeung, W-K Wong, and WC~Woo.
\newblock Convolutional lstm network: A machine learning approach for precipitation nowcasting. arxiv 2015.
\newblock \emph{arXiv preprint arXiv:1506.04214}, 2015.

\bibitem[Shor et~al.(2020)Shor, Jansen, Maor, Lang, Tuval, de~Chaumont~Quitry, Tagliasacchi, Shavitt, and Emanuel]{Shor_2020}
Joel Shor, Aren Jansen, Ronnie~Zvi Maor, Oran Lang, Omry Tuval, Félix de~Chaumont~Quitry, Marco Tagliasacchi, Ira Shavitt, and Dotan Emanuel.
\newblock Towards learning a universal non-semantic representation of speech.
\newblock 2020.
\newblock URL \url{https://www.isca-speech.org/archive/Interspeech_2020/abstracts/1242.html}.

\bibitem[Simonyan and Zisserman(2014)]{simonyan2014very}
Karen Simonyan and Andrew Zisserman.
\newblock Very deep convolutional networks for large-scale image recognition.
\newblock \emph{arXiv preprint arXiv:1409.1556}, 2014.

\bibitem[Tracey et~al.(2011)Tracey, Comina, Larson, Bravard, López, and Gilman]{6091487}
Brian~H. Tracey, Germán Comina, Sandra Larson, Marjory Bravard, José~W. López, and Robert~H. Gilman.
\newblock Cough detection algorithm for monitoring patient recovery from pulmonary tuberculosis.
\newblock In \emph{2011 Annual International Conference of the IEEE Engineering in Medicine and Biology Society}, pages 6017--6020, 2011.
\newblock \doi{10.1109/IEMBS.2011.6091487}.

\bibitem[Valk and Alumäe(2021)]{9383459}
Jörgen Valk and Tanel Alumäe.
\newblock Voxlingua107: A dataset for spoken language recognition.
\newblock In \emph{2021 IEEE Spoken Language Technology Workshop (SLT)}, pages 652--658, 2021.
\newblock \doi{10.1109/SLT48900.2021.9383459}.

\bibitem[Verma and Berger(2021)]{verma2021audio}
Prateek Verma and Jonathan Berger.
\newblock Audio transformers: Transformer architectures for large scale audio understanding. adieu convolutions.
\newblock 2021.

\bibitem[Vizza et~al.(2019)Vizza, Tradigo, Mirarchi, Bossio, Lombardo, Arabia, Quattrone, and Veltri]{VIZZA201945}
Patrizia Vizza, Giuseppe Tradigo, Domenico Mirarchi, Roberto~Bruno Bossio, Nicola Lombardo, Gennarina Arabia, Aldo Quattrone, and Pierangelo Veltri.
\newblock Methodologies of speech analysis for neurodegenerative diseases evaluation.
\newblock \emph{International Journal of Medical Informatics}, 122:\penalty0 45--54, 2019.
\newblock ISSN 1386-5056.
\newblock \doi{https://doi.org/10.1016/j.ijmedinf.2018.11.008}.
\newblock URL \url{https://www.sciencedirect.com/science/article/pii/S1386505618307639}.

\bibitem[Wyse(2017)]{wyse2017audio}
Lonce Wyse.
\newblock Audio spectrogram representations for processing with convolutional neural networks.
\newblock \emph{arXiv preprint arXiv:1706.09559}, 2017.

\bibitem[Zaman et~al.(2023)Zaman, Sah, Direkoglu, and Unoki]{zaman2023survey}
Khalid Zaman, Melike Sah, Cem Direkoglu, and Masashi Unoki.
\newblock A survey of audio classification using deep learning.
\newblock \emph{IEEE Access}, 2023.

\bibitem[Zhang et~al.(2023)Zhang, Xu, Zhang, and Tao]{zhang2023vitaev2}
Qiming Zhang, Yufei Xu, Jing Zhang, and Dacheng Tao.
\newblock Vitaev2: Vision transformer advanced by exploring inductive bias for image recognition and beyond.
\newblock \emph{International Journal of Computer Vision}, pages 1--22, 2023.

\end{thebibliography}
